\documentclass[twocolumn,aps,showpacs,  
tightenlines,superscriptaddress]{revtex4}
\usepackage{epsfig}

\begin{document}

\title{Scaling properties of 
proton-nucleus total reaction cross sections}

\author{Badawy Abu-Ibrahim}
\affiliation{RIKEN Nishina Center, RIKEN, Wako-shi, 
Saitama 351-0198, Japan}
\affiliation{Department of Physics, Cairo University, 
Giza 12613, Egypt}
\author{Akihisa Kohama}
\affiliation{RIKEN Nishina Center, RIKEN, Wako-shi, 
Saitama 351-0198, Japan}

\date{\today}

\begin{abstract}
We study the scaling properties of proton-nucleus 
total reaction cross sections for {\em stable} nuclei 
and propose an approximate expression 
in proportion to 
$Z^{2/3}\sigma_{pp}^{\rm total} + N^{2/3} \sigma_{pn}^{\rm total}$.
Based on this expression, 
we can derive a relation that enables us to predict
a total reaction cross section for any stable nucleus 
within $10\%$ uncertainty at most, 
using the empirical value of 
the total reaction cross section of a given nucleus. 
\end{abstract}


\maketitle

Recently, we studied total reaction cross sections 
of carbon isotopes ($N = 6-16$)
incident on a proton for a wide energy region \cite{bhks}. 
In that work, using our numerical results, 
we found an empirical formula, Eq.~(13) of Ref.~\cite{bhks} 
(see the Appendix). 
In this Brief Report, 
as an extension of the empirical formulae, 
we study the scaling properties of proton-nucleus 
total reaction cross sections for {\it stable} nuclei 
of the whole mass-number region. 
Our new expression is found to be in proportion to 
$Z^{2/3} \sigma_{pp}^{\rm total} + N^{2/3} \sigma_{pn}^{\rm total}$.
Although this is merely a parameterization without any 
microscopic foundation, 
once one is able to accept this expression, 
we can derive a new relation~(\ref{reac2}) 
that enables us to predict
a total reaction cross section for any stable nucleus 
using only the empirical values of the total reaction cross section 
of a given nucleus at a given energy within 10$\%$ uncertainty. 

We assume here the following expression for 
the total reaction cross sections of a proton incident on 
a nucleus with the mass number, $A (= N + Z)$:
\begin{eqnarray}  
   &{}& \sigma_{\rm R}(Z,N,E)
\nonumber\\
   &=&  \pi C(E,A) \{Z^{2/3}\sigma_{pp}^{\rm total}(E) 
                   + N^{2/3}\sigma_{pn}^{\rm total}(E) \}
\nonumber\\ 
   &\simeq& \pi C(E) A^{2/3} \times
\nonumber\\
   & & \left[ (Z/A)^{2/3}\sigma_{pp}^{\rm total}(E) 
            + (N/A)^{2/3}\sigma_{pn}^{\rm total}(E) \right],
\label{reac1} 
\end{eqnarray}
where $Z (N)$ is the number of protons (neutrons) 
in the target nucleus, 
$\sigma_{pp}^{\rm total}$ ($\sigma_{pn}^{\rm total}$) is 
the proton-proton (proton-neutron) total cross section, and 
$C(E)$($\equiv \langle C(E, A) \rangle_A$) 
is an energy-dependent coefficient to be deduced from data. 
$\langle \cdots \rangle_A$ implies the average value of
the values for different mass numbers at each energy. 
Note that Eq.~(\ref{reac1}) is merely 
a parameterization without any microscopic foundation 
and that the factorization of $C(E)$ is crucial for 
later discussion. 
This expression can predict 
the energy dependence of total cross sections of lead 
($Z = 82$, $N = 126$) with acceptable uncertainties, 
but, unfortunately, 
the isospin dependence is not necessarily adequate 
for describing neutron-rich unstable nuclei.


\begin{figure}[t]
\epsfig{file=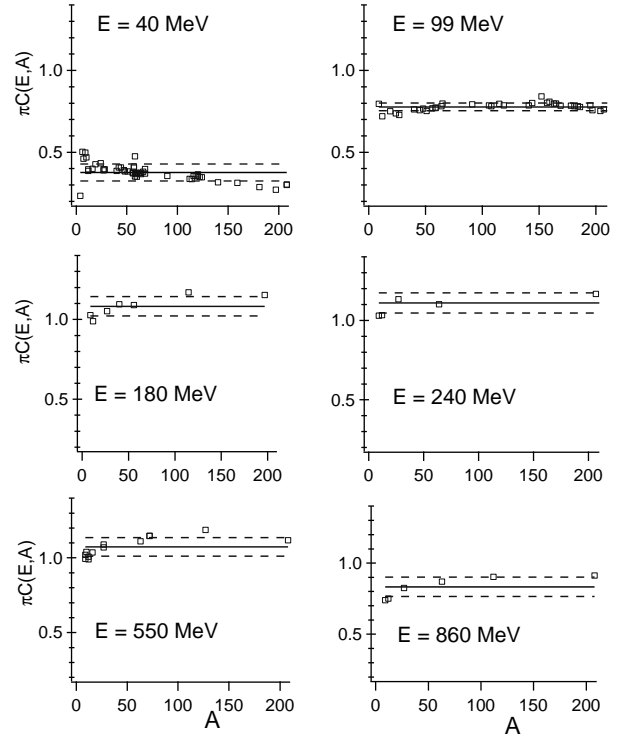,scale=0.5}
\caption{Values of $\pi C(E, A)$ 
as functions of target mass number, $A$, for each energy, $E$, 
indicated by open squares. 
For convenience, we plot $\pi C(E, A)$ instead of $C(E, A)$. 
The solid lines indicate the average values. 
The dashed lines indicate standard deviations 
around each average value. 
The experimental data are taken from Ref.~\cite{carlson}. }
\label{call}
\end{figure}

In Eq.~(\ref{reac1}), 
$C(E, A)$ in the second line is replaced by $C(E)$, 
because $C(E, A)$ depends weakly on $A$. 
Figure~\ref{call} shows the values of 
$\pi C(E, A)$ as functions of the mass 
number for all available nuclei 
at six different energies~\cite{carlson}. 
The solid lines in each panel represent 
the values of $\pi C(E)$ obtained by 
averaging $\pi C(E, A)$ over the whole mass numbers for each energy. 
The dashed lines denote 
the standard deviations of $\pi C(E)$.
These results validate the replacement 
of $C(E, A)$ with $C(E)$ in Eq.~(\ref{reac1}). 


\begin{figure}[t]
\epsfig{file=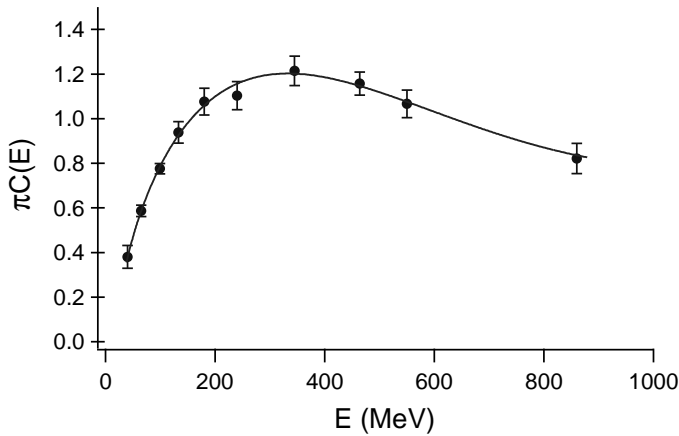,scale=0.65}
\caption{Values of coefficient, $\pi C(E)$, 
as functions of proton incident energy, $E$.
The solid curve shows the behavior of Eq.~(\ref{c-value}).
}
\label{const}
\end{figure}

\begin{table}[b]  
\caption{Values of coefficient, $\pi C(E)$, defined 
in Eq.~(\ref{reac1}), 
and their standard deviations. 
The curves in Fig.~\ref{pn1} are drawn using these values.
}
\label{ba}
\begin{center}
\begin{tabular}{cc||cc}
\hline\hline
$E$ (MeV) &  $\pi C(E)$ & $E$ (MeV) &  $\pi C(E)$ \\
\hline
40  & 0.3764$\pm$0.051 & 240 & 1.1104$\pm$0.063 \\
65  & 0.5868$\pm$0.025 & 345 & 1.2144$\pm$0.066 \\
99  & 0.7776$\pm$0.023 & 464 & 1.1570$\pm$0.052 \\
133 & 0.9384$\pm$0.048 & 550 & 1.0741$\pm$0.062 \\
180 & 1.0822$\pm$0.060 & 860 & 0.8334$\pm$0.068 \\
\hline\hline
\end{tabular}
\end{center}
\end{table}

In Fig.~\ref{const}, we summarize the values of $\pi C(E)$ as 
a function of energy. 
Those values are also listed in 
Table~\ref{ba} with their standard deviations defined by
\begin{equation}
   \Delta C(E) = \sqrt{{1 \over N_{\rm data}}
                 \sum_{i = 1}^{N_{\rm data}} (C_i - \bar{C})^{2} }.
\label{dev}                 
\end{equation}

To facilitate the numerical calculations of $\sigma_{\rm R}$ using 
Eq.~(\ref{reac1}), 
it would be useful to prepare the analytical form for $C(E)$ 
as a function of proton incident energy, $E$. 
We adopt, for example,    
\begin{eqnarray}
  \pi C(E)&=&a_{1}-a_{2} \,{\rm exp}(-a_{3}E^{a_{4}})\nonumber\\
          &\times&{\rm cos}[a_{5}E^{a_{6}}], 
\label{c-value}
\end{eqnarray} 
where $E$ is the projectile kinetic energy 
in units of MeV.
The constants are given by  
$a_{1}$=0.909824,  $a_{2}$=2.329619, 
$a_{3}$=0.4345765, $a_{4}$=0.2655287,  
$a_{5}$=0.075, and $a_{6}$=0.6262386.
The behavior of Eq.~(\ref{c-value}) is represented 
by the solid curve in Fig.~\ref{const}.


\begin{figure}[t]
\epsfig{file=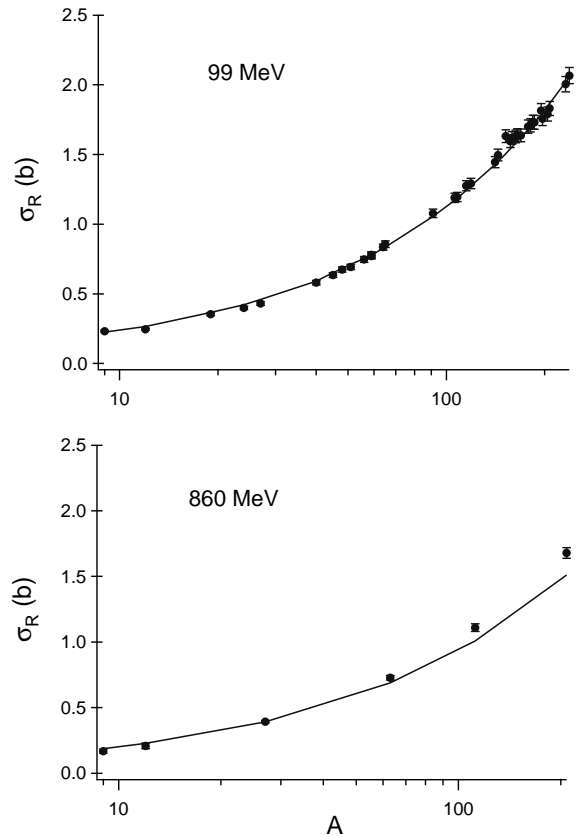,scale=0.53}
\caption{Comparison of the numerical results obtained using 
Eq.(\ref{reac1}) with the empirical data of 
total reaction cross sections for proton-nucleus 
reactions as a function of mass number, $A$, 
for $E$ $= 99$ and $860$ MeV. 
The solid curves are the numerical results. 
The values of $\pi C(E)$ are listed in Table~\ref{ba}. 
The experimental data are taken from Ref.~\cite{carlson}. 
}
\label{pn1}
\end{figure}

Using the preceding setups, in Fig.~\ref{pn1} 
we show the results of the formula with the data 
of $\sigma_{\rm R}$ as a function of mass number, $A$, 
at two different energies. 
The solid curves show the results of Eq.~(\ref{reac1}). 
The values of the coefficient, $\pi C(E)$, 
that provide the solid curves 
are listed in Table~\ref{ba}. 
The agreement of the numerical results with the empirical data 
is fairly good. 
Although we do not show the results for $E=40$ MeV, 
one should be careful in the application 
of this formula at $E=40$ MeV because of the large deviation 
from the data for $A \ge 100$.

\begin{figure}[t]
\epsfig{file=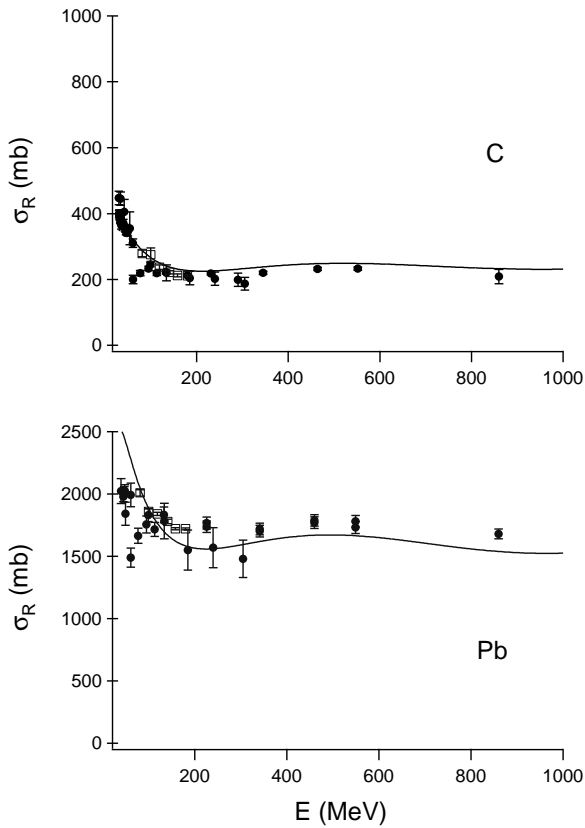,scale=0.53}
\caption{Comparison of the numerical results obtained using 
Eqs.(\ref{reac1}) and (\ref{c-value}) with the empirical data of 
the total reaction cross sections for proton-nucleus 
reactions as a function of energy, $E$. 
The solid curves are the numerical results 
of Eqs.~(\ref{reac1}) and (\ref{c-value}).  
The experimental data are 
taken from Ref.~\cite{carlson} (solid circles) and 
Ref.~\cite{Auce} (open squares). 
}
\label{RCS_E}
\end{figure}

\begin{figure}[t]
\epsfig{file=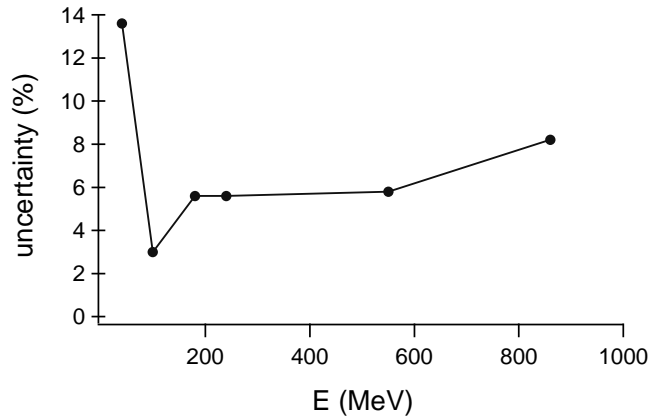,scale=0.65} 
\caption{Uncertainties in numerical results 
of $\sigma_{\rm R}$ 
of Eq.~(\ref{reac1}) as functions of proton incident energy, $E$. 
The solid line is a guide for the eye.
}
\label{uncertainty}
\end{figure} 

Also, we show the results of the formula with the empirical data 
of $\sigma_{\rm R}$ as a function of proton incident energies 
in Fig.~\ref{RCS_E}. 
The solid curves represent 
the numerical results of Eqs.~(\ref{reac1}) and (\ref{c-value}) 
for the total reaction 
cross section of protons incident on two target nuclei, such as 
C and Pb.

To show the precision of our formula, Eq.~(\ref{reac1}), 
we estimate the fluctuations 
of the data around the numerical results at each energy. 
We introduce the fluctuation defined by 
the standard deviations of $C(E)$ 
divided by its mean values,
that is, $\Delta C(E)/C(E)$, 
and plot them as a function of energy, $E$, 
in Fig.~\ref{uncertainty}. 
One can see that, except for the case of 40 MeV, 
the uncertainties in our numerical results are less than 10$\%$. 
That is, the feature of each nucleus 
affects the magnitude of the total reaction cross section 
by 10$\%$ at most.


Let us derive the ``parameter-free" expression 
of $\sigma_{\rm R}$ using Eq.~(\ref{reac1}), 
which is the major topic of this article. 
In Eq.~(\ref{reac1}), we introduced 
the coefficient, $C(E)$ 
as being independent of $A$ within 10$\%$ uncertainty.
If we accept this fact, we can estimate 
a total reaction cross section of a nucleus $A_{1} (=Z_{1}+N_{1})$ 
at a given energy using a known value of the total 
reaction cross section of another nucleus $A_{2} (=Z_{2}+N_{2})$ 
at the same energy. 
The expression becomes 
\begin{equation}  
\sigma_{\rm R}(Z_{1},N_{1})
  = (1 \pm \Delta)
    \left(\frac{\sigma_{pp}Z_{1}^{2/3}+\sigma_{pn}N_{1}^{2/3}}
               {\sigma_{pp}Z_{2}^{2/3}+\sigma_{pn}N_{2}^{2/3}}
    \right)
    \sigma_{\rm R}(Z_{2},N_{2}), 
\nonumber\\
\label{reac2} 
\end{equation}
where $\Delta$ implies the uncertainty coming from 
that of $C(E)$. 
Here, $\Delta$ is defined by 
$\Delta$$\equiv \langle\delta(E)\rangle_E$, 
and $\delta(E)$$\equiv 2\Delta C(E)/C(E)$. 
$\langle\delta(E)\rangle_E$ implies 
the average of $\delta(E)$ values at various energies. 
We obtain $\Delta \simeq 0.1$.
No fitting parameter appears here, 
but, at most, 10$\%$ uncertainty should be taken into account.

\begin{figure}[t]
\epsfig{file=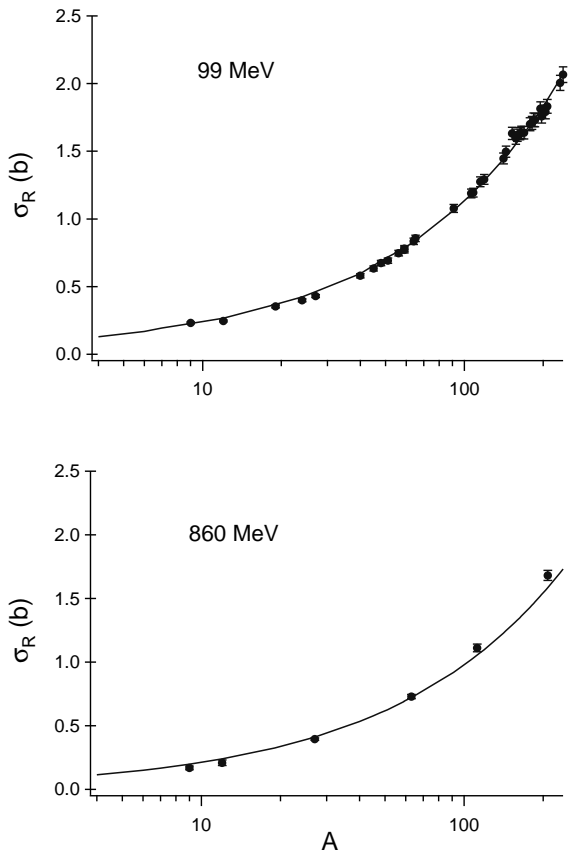,scale=0.53} 
\caption{Comparison of the numerical results of Eq.~(\ref{reac2}) 
with the empirical data of 
the total reaction cross sections for proton-nucleus reactions 
as functions of mass number, $A$. 
The solid curves represent the numerical results 
obtained using Eq.~(\ref{reac2}). 
The experimental data are taken from Ref.~\cite{carlson}. 
}
\label{pn3}
\end{figure}

In Fig.~\ref{pn3}, 
we show the numerical results of Eq.~(\ref{reac2}) 
with the empirical data as a function of mass number, $A$, 
at 99 and 860 MeV. 
For the reference nucleus, 
which corresponds to $A_2$ in Eq.~(\ref{reac2}), 
we here choose a nucleus of a medium mass number, 
such as $^{56}$Fe, $^{64}$Zn, and $^{64}$Cu. 
For numerical calculations of Eq.~(\ref{reac2}) 
for 40, 99, 180, 550 and 860 MeV, 
we choose a nucleus and the reaction cross section 
at each energy as 
($^{64}$Zn, 1092$\pm$23 mb),   
($^{64}$Cu, 835$\pm$23 mb),
($^{56}$Fe, 662$\pm$19 mb),
($^{64}$Cu, 667$\pm$67 mb),
($^{64}$Cu, 777$\pm$17 mb) and
($^{64}$Cu, 728$\pm$17 mb), respectively~\cite{carlson}.
Here we neglect $\Delta$. 

The expression~(\ref{reac2}) indicates that, 
if we have the empirical value of 
the reaction cross section of a given nucleus, 
we can predict the reaction cross section of other nuclei.
The numerical results are consistent with the empirical data 
of stable nuclei. 
A possible way of estimating an unknown reaction cross section 
of a given nucleus at a given energy 
is to use Eqs.~(\ref{reac1}) and (\ref{c-value}) 
or to use Eq.~(\ref{reac2}). 
We do not need any parameter for the latter. 
The fluctuations in $\sigma_{\rm R}$ 
coming from the feature of each nucleus 
affects the magnitude by 10$\%$ at most.

In summary, we have introduced an approximate 
but new expression for  
proton-nucleus total reaction cross sections 
in terms of the number of protons and neutrons, 
and the total cross sections of proton-proton and proton-neutron 
reactions.
This expression is applicable in the energy 
range from 40 MeV to 1000 MeV.
Also, we have derived a simple relation 
for predicting the total reaction cross sections 
of a proton with any nucleus within 10$\%$ uncertainty,
using the empirical values of the total reaction cross section of 
a given nucleus at a given energy. 
A better result is obtained if a neighboring nucleus is adopted.

\vspace{5mm}
We acknowledge T. Nakatsukasa for his comments and 
Y. Suzuki for carefully reading the manuscript. 
A.K. acknowledges K. Iida and K. Oyamatsu 
for their helpful discussion and comments. 
A part of this work is supported by 
the Japan International Cultural Exchange Foundation (JICEF).

\vspace{10mm}

\appendix

\section{Volume-type form}

\begin{figure}[t]
\epsfig{file=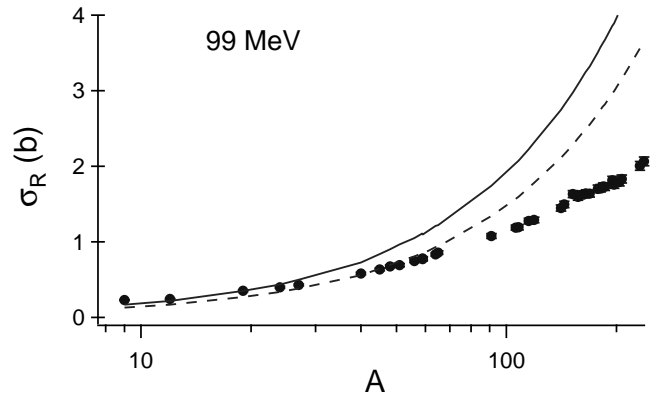,scale=0.65} 
\caption{Prediction using Eq.~(\ref{reac-vol}) 
for the proton-nucleus total reaction cross sections at 99 MeV. 
The nuclei with 4 $<$ $A$ $<$ 30 fit the results 
with $C(E)=0.3523198$ 
(solid curve) 
and those with 30 $<$ $A$ $<$ 60 
fit the results with $C(E)= 0.27075$ (dashed curve).
}
\label{vol}
\end{figure}

We found that, in Eq.~(13) of Ref.~\cite{bhks}, 
for all the carbon isotopes, 
the following relation is satisfied over the entire energy range: 
\begin{equation}  
   \frac{\sigma_{\rm R}(p + {_{\quad 6}^{6\!+\!N}{\rm C}})}
        {\sigma_{\rm R}(p + {^{12}_{\ \, 6}{\rm C}})}
   = R({\rm C}) \frac{6 \sigma_{pp}^{\rm total} 
                    + N \sigma_{pn}^{\rm total}}
                     {6 \sigma_{pp}^{\rm total} 
                    + 6 \sigma_{pn}^{\rm total}},
\label{emp} 
\end{equation}
\vspace{25mm}
where $R({\rm C}) = 0.96 \pm 0.05$ and $N \ge 7$.
The above relation suggests 
the following volume-type formula:  
\begin{equation}  
   \sigma_{\rm R}(Z, N, E)
   = C(E) \left[Z \sigma_{pp}^{\rm total}(E) 
              + N \sigma_{pn}^{\rm total}(E) \right].
\label{reac-vol} 
\end{equation}

We find that this formula is able 
to estimate the proton-nucleus cross section 
in the mass number range around $b<$ $A$ $< b+30$, 
where $b$ is any number. 
The applicability is limited to a certain restricted region in $A$. 
Figure~\ref{vol} shows the numerical results 
of Eq.~(\ref{reac-vol}).
The nuclei with 4 $<$ $A$ $<$ 30 fit the results 
with $C(E)=0.3523198$ 
(solid curve) 
and those with 30 $<$ $A$ $<$ 60 
fit the results with $C(E)= 0.27075$ (dashed curve).


\vspace{10mm}



\begin{thebibliography}{99}
\bibitem{bhks} B. Abu-Ibrahim, W. Horiuchi, A. Kohama and Y. Suzuki, 
        Phys. Rev. C {\bf 77}, 034607 (2008); 
         Erratum: {\it ibid}. C {\bf 80}, 029903 (2009). 
\bibitem{carlson} R. F. Carlson, 
        At. Data and Nucl. Data Tables {\bf 63}, 93 (1996). 
\bibitem{Auce} A. Auce, A. Ingemarsson, R. Johansson, 
        M. Lantz {\it et al.}, 
        Phys. Rev. C {\bf 71}, 064606 (2005).
\end{thebibliography}
\end{document}